# Temporal Load Imbalance on Ondes3D Seismic Simulator for Different Multicore Architectures

Ana Luisa Veroneze Solórzano, Philippe Olivier Alexandre Navaux, Lucas Mello Schnorr

Informatics Institute (PPGC/UFRGS), Porto Alegre, Brazil

*Abstract*—The variety of today's multicore architectures motivates researchers to explore parallel scientific applications on different platforms. Load imbalance is one performance issue that can prejudice parallel applications from exploiting the computational power of these platforms. Ondes3D is a scientific application for seismic waves simulation used to assess the geological impact of earthquakes. Its parallelism relies on applying a regular domain decomposition in the geological domain provided and distributing each sub-domain to MPI ranks. Previous works investigate the Ondes3D significant spatial and temporal imbalance and suggest new parallelization and load balance techniques to minimize them. However, none explored its execution on different architectures. Our paper evaluates the Ondes3D performance for two earthquake scenarios in eight different multicore architectures, including Intel, AMD, and ARM processors. We measure the load distribution per MPI rank, evaluate the temporal load imbalance and compare the application's kernels executions. Our results show that the temporal load imbalance of Ondes3D depends on the architecture choice, presenting the best platform to minimize such imbalance.

## I. INTRODUCTION

High-Performance Computing community is making an effort to provide grids and clusters for scientific research. Access to platforms highly configurable, controllable, and monitored, with minimal infrastructure management, motivates research advance [1]. With multiple generations of CPU architectures available, researchers can exploit the performance of scientific applications running on different experimental setups to find a better choice.

Proper load distribution is key to harness the power of multicore processors. Uneven load distribution across the resources (e. g., processes, processors, nodes) generates idle states where a resource waits for delayed ones [2]. Domain decomposition is the first step to distribute data into parallel resources [3]. Irregular decomposition requires extra computations to decide the different subdomains' sizes. Regular decomposition into subdomains of equal size facilitates the load distribution but fails to guarantee the same amount of load per subdomain.

Ondes3D is a scientific application used for simulating the seismological motion of earthquake scenarios [4]. The code is parallelized with MPI and presents a spatial and temporal load imbalance [5]. Previous works suggest methods to minimize the Ondes3D imbalance by using a parallelization approach for the numerical discretizations using MPI and OpenMP [6], and a dynamic rebalancing scheme using Adaptive MPI on top of the Charm++ runtime [7]. However, the studies explored a single processor, leaving uncovered the investigation of the temporal and spatial imbalance in different architectures. One hypothesis for exploring a varied set of multicore processors is to see if the imbalance increase in a specific environment. This effort would minimize the researchers' work to improve the application performance by only choosing the proper platform.

In this paper, we characterize the temporal imbalance of Ondes3D on eight different microarchitectures for two earthquake scenarios. We conducted an analysis of the applications' computational signature using trace files collected at runtime. We evaluate the global load distribution for each scenario, the temporal imbalance, and the load distribution considering the application's kernels. In summary, our contributions are:

- Analyze the Ondes3D performance for two earthquake scenarios in eight different multicore architectures;
- Evaluate the load distribution per MPI rank;
- Evaluate the temporal imbalance for each scenario and compare the executions per architecture;
- Evaluate the kernels executions individually, which are responsible for the main computation;
- Demonstrate that the Ondes3D temporal load imbalance depends on the architecture choice;
- Indicate the best architecture to minimize such imbalance;

This paper is organized as follows. Section II presents the Ondes3D application. Section III presents our methodology, with details of the software and hardware used. Section IV presents our findings. Section V presents other studies that cover performance analysis in different architectures. Section VI finishes the paper with a discussion and conclusion.

## II. ONDES3D: SEISMIC WAVE SIMULATOR

Ondes3D is a seismic wave propagation simulator developed by the French Geological Survey (BRGM) that can be used for earthquake modeling and other strong ground motion evaluations. The application implements the Finite-Difference Method (FDM) to model seismological motion and applies the Convolutional Perfectly Matched Layer (CPML), to determine a boundary region from an unbounded one, by absorbing the energy in the artificial limits [8], [9].

The application parallelism relies on the 3D domain decomposition into subdomains of equal size processed by different MPI ranks [5]. Ondes3D receives an input file characterizing the geological domain and another to determine the N × M topology of the decomposition. The number of MPI ranks must match the topology determined. Figure 1a illustrates a domain partitioning for a 4 × 4 topology (A), that would run in 16 MPI

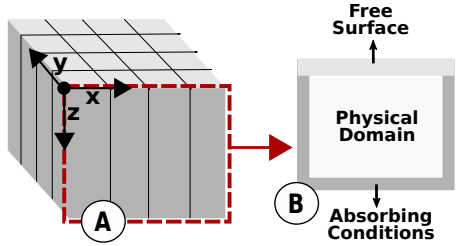

(a) The 3D domain partition for a 4 x 4 topology (A), and a lateral view of the three domain regions (B).

(b) The main loop of Ondes3D (left). Each iteration call four macro kernels that launch microkernels (right).

Fig. 1: Ondes3D application: (a) the domain decomposition and the three domain regions; (b) the main loop of the application.

ranks. The code implements a bidimensional decomposition, considering only two of the three cartesian coordinates, x and y. Each subdomain computation depends on the region where the cells are located, as presented in a lateral view of the domain in Figure 1a (B): cells at the Physical Domain will compute mostly FDM calculations, cells at the bottom and lateral borders will compute Absorbing Boundary Conditions using CPML, and cells at the boundary top will suffer from computations related to the Free Surface conditions [9].

Figure 1b presents the main loop of the application, where each iteration works over a subdomain. It computes well-defined iterations interleaved with communications that represent significant time intervals for the simulation. The iterations call four macro kernels that perform frequent asynchronous communications among MPI ranks: seisMoment, intermediates, stress, and velocity. Three kernels call different microkernels based on the domain region and the geological characteristics of the cell. Even with subdomains of equal size, the parallel ranks compute different loads.

Ondes3D suffers from spatial and temporal load imbalance. The spatial imbalance comes from the rock layer's geological characteristics and the execution of the computational methods for processing the wave propagation [10]. Cells located at the Absorbing Boundary Conditions compute more load than cells in the inner part of the region, because of the extra absorbing computations [6]. The temporal load imbalance comes from the load variation between execution times of different iterations. Previous work reports that the temporal imbalance is more substantial than the spatial one due to low-level optimizations in CPU when processing the microkernels, which results in more cycles for some iterations [5].

III. MULTICORE ARCHITECTURES AND METHODOLOGY

This section presents the multicore architectures used to execute Ondes3D and our methodology towards the temporal imbalance analysis. The platforms are hosted in the High-Performance Computational Park (http://gppd-hpc.inf.ufrgs.br/), and in the Grid'5000 [11] (http://grid5000.fr), a testbed for research on parallel and distributed systems.

We explore platforms from different vendors and families to evaluate Ondes3D performance considering the architectures' evolution. Table I shows the characteristics of the eight multicore architectures chosen. The number of cores represents the total cores available per node. It is ordered by the oldest (Nehalem) to the most modern architecture (Zen 2). We selected two Skylake processors, a recent architecture that we had access to, so we will reference them as Skylake6126 and Skylake6130. Concerning the network interconnection, we have 100Mbps Ethernet for Nehalem, Ivy Bridge, and Haswell; 2×25Gbps Ethernet for Skylake6130 and Zen; 10Gbps Ethernet and 100Gbps Omni-Path for Skylake6126; and 10Gbps with 100Gbps Ethernet for ThunderX2. All nodes have Debian 4.19.160-2 as operating system.

TABLE I: Processor's characteristics.

| Year | Architecture | Processor/Node | Cores | Speed |
|---|---|---|---|---|
| 2008 | Nehalem | 4 × Intel Xeon X7550 | 32 | 2,0GHz |
| 2012 | Ivy Bridge | 2 × Intel Xeon E5-2640v2 | 16 | 2,0GHz |
| 2013 | Haswell | 2 × Intel Xeon E5-2650v3 | 20 | 2,3GHz |
| 2014 | Broadwell | 1 × Intel Xeon E5-2620v4 | 8 | 2,1GHz |
| 2015 | Skylake | 2 × Intel Xeon Gold 6126 | 24 | 2,6GHz |
| 2015 | Skylake | 4 × Intel Xeon Gold 6130 | 64 | 2,1GHz |
| 2017 | Zen | 2 × AMD EPYC 7301 | 32 | 2,2GHz |
| 2018 | ThunderX2 | 2 × ARM ThunderX2 99xx | 64 | 2,5GHz |
| 2019 | Zen 2 | 1 × AMD Ryzen 9 3950X | 16 | 3,5GHz |

We consider the maximum number of cores available for the Zen 2 architecture to define the domain decomposition. We choose a topology of 4 × 4 (16 MPI processes) that presents sufficient subdomains to identify the Ondes3 spatial and temporal imbalance. All setups are single-node except for the Broadwell architecture, in which we use two nodes with one processor each. We only evaluate the time spent computing load without the time communicating between processes.

We select two earthquake scenarios: Chuetsu-Oki, which happened in 2007 in Japan, and Sichuan [12], which occurred in 2008 at China [13]. Chuetsu-Oki is a more simple input, with one geological layer to be computed over 500 iterations. Sichuan is a more complex input and is also more time and memory costly with four layers, resulting in more cells to be computed over 1000 iterations.

Figure 2 depicts our methodology workflow, defined in three steps. We select a single representative run for each architecture and case. (i) In the *Pre computation* step, we define the experimental project by selecting and configuring the platforms. We use the Spack package manager to install the software required in the different platforms [14], we define the experimental plan by selecting the workloads and user-configurable parameters, and we do manual instrumentation in the Ondes3D code with Score-P. (ii) During the *Computation*, we run the application compiled against OpenMPI version 3.1.6, GCC version 8.3.0 using the flag *-O3*, and Score-P version 6.0. (iii) After collecting the trace files, we start the *Post computation*. Score-P generates traces in Open Trace Format (OTF2) converted to CSV using otf2utils (https://github.com/schnorr/otf2utils). We convert the CSV file to the `feather` format and filter the data to obtain: the MPI ranks number, iterations (500 for Chuetsu-Oki and 1000 for Sichuan), operations (seisMoment, stress, intermediates, and velocity), and the computational load, calculated with the start and end time of the operations. Finally, we write the data manipulation in the R programming language, using tidyverse [15], a collection of R packages for data science, to create the visualizations for our analysis.

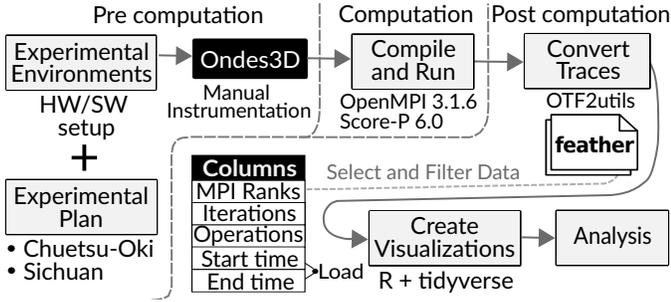

Fig. 2: The methodology workflow for our evaluation.

## IV. CHARACTERIZING ONDES3D LOAD IMBALANCE

This section presents our results of the Ondes3D temporal imbalance for the experimental environments selected (see the previous section). We assume the *computational load* as the time spent computing the four macro kernels (seisMoment, stress, intermediates, and velocity). Section IV-A presents a global overview of the load distribution per processor and MPI ranks. Section IV-B presents a temporal load imbalance investigation, characterizing the imbalance throughout the iterations. Section IV-C presents the temporal imbalance considering the macro kernels. Section IV-D presents an additional investigation of the microkernels.

### A. Global Load Distribution

Figure 3 presents the execution times for the 16 MPI ranks running the Chuetsu-Oki scenario (top facets) and the Sichuan (bottom). Each facet represents an architecture. All architectures present an execution time variation per rank for both scenarios. The load is also varied when comparing the processors, for example, rank 15 of the Zen 2 computes the highest load for Chuetsu-Oki, which was 4.9 times smaller than the most loaded rank for the ThunderX2 (annotation A). Skylake6126 and Skylake6130 present similar execution times for both workloads (B). Chuetsu-Oki computes similarly at Zen, Haswell, and Broadwell (C1), while Sichuan computes faster at Zen than at Haswell and Broadwell (C2), indicating that more complex inputs perform better at Zen than in the other two Intel processors.

Going to some specifics of Figure 3, the two most time-consuming architectures for both cases are Nehalem and ThunderX2. ThunderX2 has a more significant load variation than Nehalem for Chuetsu-Oki. Nehalem time difference from the most loaded rank to the less loaded is 247 seconds, and for ThunderX2, the value is 511s. For Sichuan, rank 7 computes the highest load at Nehalem in 29733s (D1), and rank 12 at ThunderX2 in 24835s (D2). Ranks 7 and 12 compute the domain's borders, which is more costly than the inner part due to the Absorbing Conditions. Zen 2 presents the best performance in both scenarios achieving a maximum load of 6755s in rank 7 (E). Sichuan shows a less diverse load distribution per rank for most architectures than Chuetsu-Oki, probably because Chuetsu-Oki execution is less costly with only one geological layer and Sichuan with four layers. The ARM ThunderX2 benefits from this characteristic, performing Sichuan faster than Nehalem, while the Intel and AMD architectures keep a similar behavior in both scenarios.

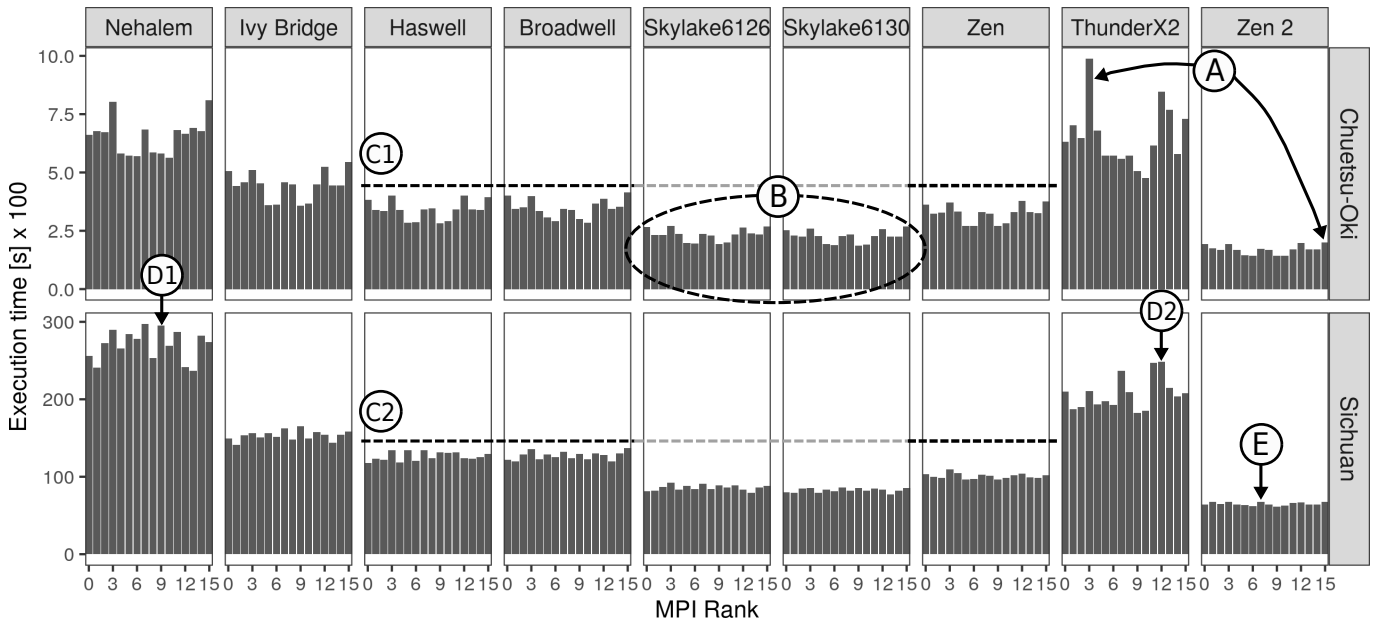

Fig. 3: Execution time for each architecture for the Chuetsu-Oki case (top facets) and Sichuan (bottom) over 16 MPI ranks.

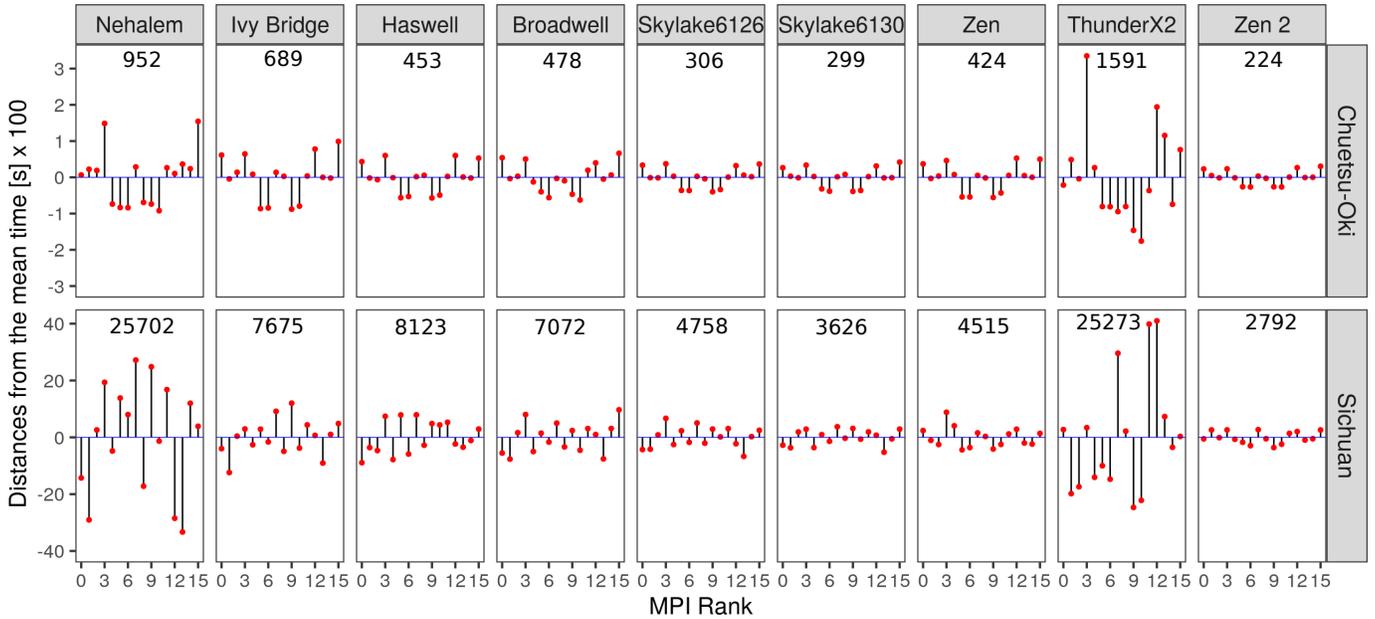

Fig. 4: Computational load distances from the mean load per MPI rank for each platform.

Figure 4 shows the computational load distances per rank from the mean load of the Chuetsu-Oki executions (top facets) and the Sichuan executions (bottom facets). Each facet represents an architecture. The number on the top is the total distance from the ideal load balance, and the horizontal line is the ideal load distribution when ranks compute the same load. Ranks are considered *overloaded* when computing more load than the mean and *underloaded* when computing less. Considering the total distances, Zen 2 presents the best result with ranks closer to the mean for both scenarios, indicating a less diverse load distribution. Nehalem and ThunderX2 makespan are less than 200s apart for Chuetsu-Oki, as seen in Figure 3, while in Figure 4 the total distance between ThunderX2 and Nehalem is more than 600s. For Sichuan, the Ivy Bridge architecture computes more load than the Haswell, as seen in Figure 3; however, the total distance from an ideal imbalance is bigger at Haswell than at Ivy Bridge.

**For Chuetsu-Oki**, Figure 4 shows a strong imbalance at ThunderX2, with rank 3 more than 300s overloaded. Nehalem imbalance was less severe, with rank 15 achieving 154s apart from the mean, and more overloaded ranks than underloaded when comparing with ThunderX2. Ranks 3 and 15 process corners of the domain. In Figure 3, we identify a similar load distribution for Haswell, Broadwell, and Zen. In Figure 4 they present the same most overloaded and most underloaded ranks but a diverse load distribution between other ranks, as we see with the total distance values. Zen 2 approximates most of its ideal load distribution, even though it still presents a variation among ranks.

**For Sichuan**, Figure 4 shows that Nehalem has an even more varied load distribution among ranks than for Chuetsu-Oki. ThunderX2 presents the most distant rank from the mean, rank 12 is more than 4000s overloaded, and half of the processes are underloaded. The opposite happens for Nehalem, where rank 12 is the most underloaded. Broadwell, Ivy Bridge, Skylake6126, and Skylake6130 present similar load distribution. We observe how balanced the AMD architectures are compared with the Intel ones until Broadwell. Zen computes Sichuan with less imbalance than Chuetsu-Oki after rank 9, up to 53s overloaded for rank 12.

### B. Temporal Load Imbalance

Our next approach was to analyze the temporal load imbalance for the 500 iterations of Chuetsu-Oki and the 1000 iterations of Sichuan. Figure 5a shows the average load processed per architecture (colored lines) during the Chuetsu-Oki (left facet) and the Sichuan (right facet) iterations. The average represents the mean load computed by all ranks per iteration and architecture. Figure 5b presents the load per rank and iterations for some architectures. We keep the vertical axis with free scales so we can notice the imbalance, even if subtle.

**For Chuetsu-Oki**, we observe in Figure 5a that all architectures, except ThunderX2 and Zen 2, present a temporal imbalance between iteration 100 and 150 (A). Nehalem achieves the highest average load of 1.49s in this interval. ThunderX2 presents modest temporal load imbalance, with slight load variation over iterations. Both Skylake processors have similar temporal imbalances as expected from processors from the same family. Zen presents more similarities with Haswell and

Broadwell, instead of with its following architecture, Zen 2. In the ThunderX2 facet on the left part of Figure 5b, we see slight temporal imbalances for all ranks throughout iterations. We can also notice the severity of the spatial imbalance, with rank 3 computing two times slower than rank 10. Below, in the Zen 2 facet, we identify the same imbalance increase presented for most processors in Figure 5a after iteration 100 (A). We also identify some outliers, mostly for rank 15 (B).

**For Sichuan**, we see in Figure 5a more stability in the load variation throughout iterations, except for Nehalem, which presents an average load increase from iteration 90 until the end of the execution (B). This increase is also noticeable for Ivy Bridge, Broadwell, Haswell, Skylake6126, and Skylake6130. Broadwell and Haswell present similar temporal imbalances. Zen is more similar to Zen 2 for Sichuan, a more complex workflow. Going to further analysis in the right part of Figure 5b, we identify an expected imbalance variability because of the Sichuan complexity with four layers. The Intel processors present similar temporal imbalance, but Haswell and both Skylake present more load variability along iterations, as seen in the Skylake6130 facet. ThunderX2 has a drastic difference in the loads computed per rank, with some points of load increases for rank 12, for example (C). Zen presents interesting points: rank 3 computes more load than others, followed by rank 4, that showed a load drop around iteration 125 (D) and then several load variation until the execution end. Zen 2 facet presents the most intense load variability, which has few interferences on the final execution time since the variation goes from 6s to 6.75s.

*C. Temporal Imbalance for the Ondes3D Macro Kernels*

Finally, we investigate the temporal imbalance considering the Ondes3D macro kernels: `intermediates`, `stress`, and `velocity`. The `seisMoment` kernel defines the workload's seismic moment without calling microkernels, computing small loads without temporal imbalance. Figure 6 presents results for the Chuetsu-Oki scenario and Figure 7 for Sichuan. They illustrate the domain decomposition for the 4 × 4 MPI processes, so each facet represents one MPI rank. This gave us an upper view of the domain, so we can identify the ranks on the borders that process more Absorbing Conditions.

**For Chuetsu-Oki**, Figure 6a shows that the temporal imbalance increase after iteration 100 is mostly due to the borders computation for the three kernels, as we can see for Skylake6130 (annotation A), Zen, and Zen 2. The Zen 2 architecture presents a more subtle imbalance. The Zen computes similar to the Skylake6130 and the other Intel architectures. Comparing the kernels' computation per architecture, the rank of coordinates [3,3] for ThunderX2 (B1) computes *stress* faster than *velocity*, and the opposite at Zen 2 (B2). Also, rank at coordinates [1,2] of ThunderX2 computes almost two times more load for *velocity* than *stress* (C), different from the other architectures. The ThunderX2 temporal load variability is present in all kernels. We only notice outliers for processing *intermediates* at the Zen 2 (D) and the ThunderX2 when analyzed at higher scales. Figure 6b shows the ranks

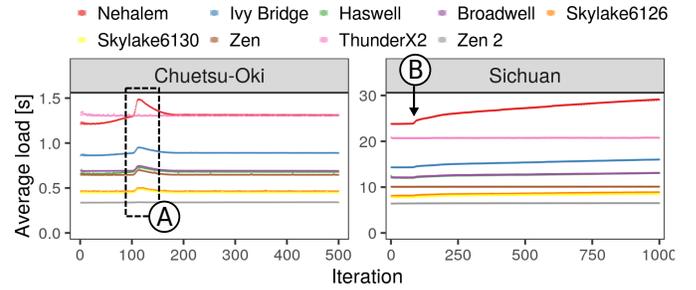

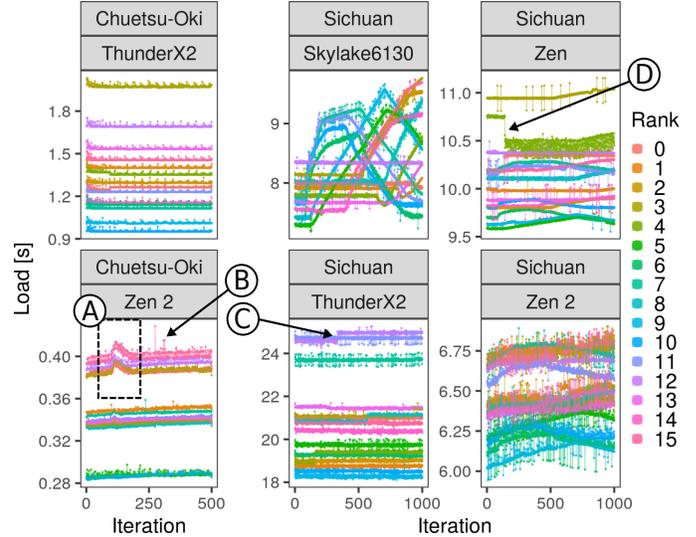

(a) Average loads of Chuetsu-Oki (left) and Sichuan (right).

(b) Load per rank for some platforms of each scenario.

Fig. 5: Temporal imbalance throughout Ondes3D iterations.

that compute mainly the Physical Domain of the geological region for Intel architectures. The *intermediates* computes a similar load for all platforms. Ivy Bridge and Haswell present a load increase for `stress` around iteration 100 (A), which disappears for Broadwell and the following architectures. Broadwell also presents less imbalance variability for *stress* and *velocity*.

**For Sichuan**, as depicts in Figure 7a, the temporal load imbalance per macro kernel presents a different behavior than Chuetsu-Oki. The load distribution is similar for the Intel architectures, as represented in the Haswell plot. The `stress` and `velocity` computation increase around iteration 100 in all Intel architectures for the ranks in the middle generating a wave format (A). We were expecting that the ranks in the middle that compute more Physical Domain operations compute more load than the ranks on the borders, which processes more Absorbing Conditions. However, they are so costly for Sichuan as computing the borders or even more costly, as seen in Zen rank at position [2,2] (B). Zen approximates to Zen 2 behavior for almost all execution with some differences to compute `stress` and `velocity` (C1 and C2). Zen, ThunderX2, and Zen 2 present fewer temporal imbalance than other architectures, which indicates that the imbalance

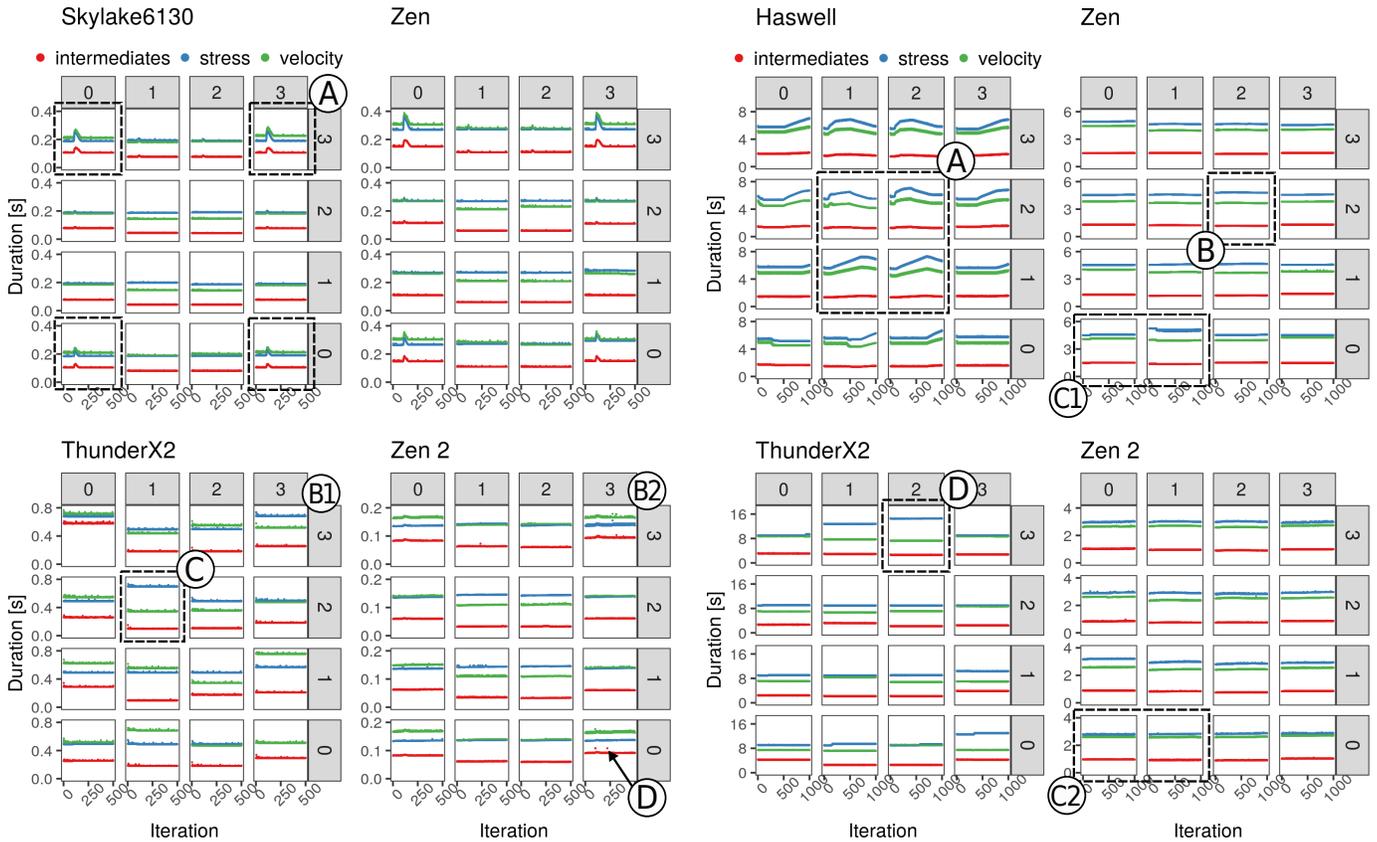

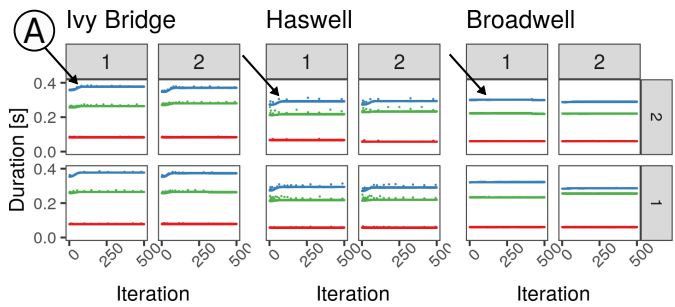

(b) The four ranks that processed the central region of Ondes3D.

Fig. 6: Load distribution for the Chuetsu-Oki macro kernels.

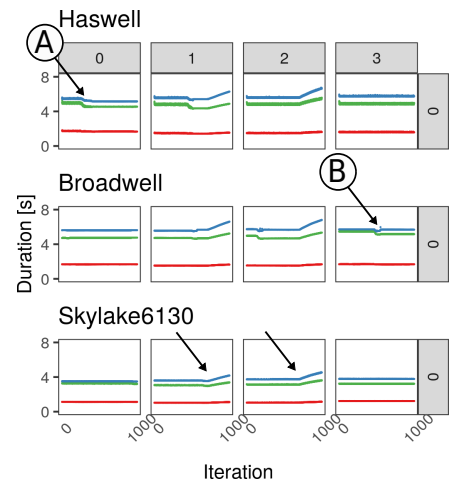

(b) Load distribution for three Intel architectures.

Fig. 7: Load distribution for the Sichuan macro kernels.

from the diverse rock's characteristics becomes minimized in these platforms. ThunderX2 presents far more time than other architectures to compute `stress` than `intermediates` at rank [2,3] (D). Figure 7b characterizes particularities on the temporal imbalance for three Intel architectures. The Haswell first rank at coordinates [0,0] presents a load decrease around iteration 200 (A), and rank [3,0] presents temporal load balance. Broadwell, the following architecture, presents the opposite, with an imbalance decrease for rank [3,1] (B) and stability for the first facet. The most recent Intel architecture evaluated, Skylake, shows temporal load stability for the first and last kernels, and the temporal imbalance in the middle is still present but minimized.

### D. Microkernels investigation

We conduct an analysis comparing the Ondes3D Assembly (ASM) source files generated in each platform. The Intel and the AMD processors generate equal files represented by x86 instructions. The same binary file generated in one of those machines executes normally in the others. The ARM processors present different instructions for all files.

We compare the ASM codes of each microkernel presented in Figure 1b to verify if their computation differs. We use the ARM ThunderX2, the AMD Zen 2, and the Intel Skylake 6130. Again, we identify several differences in the ARM's ASM code comparing with Intel and AMD. Previous work shows that adaptations on the ASM code can impact the application performance, not possible when making repairs at the source-code statement-level only [16].

To verify if those differences impact the execution's performance, we instrumented the Ondes3D code to gather the `CPML4` microkernel signature, the most called microkernel. It is launched in three different moments for *velocity* to process absorbing conditions in the subdomain, related with cells located at the x, y, and z coordinate positions. Table II presents the most loaded ranks for one iteration of Chuetsu-Oki using 16 MPI ranks. We obtain the times over 30 executions for each of the three CPML4 appearances.

TABLE II: CPML4 execution for the stress macro kernel.

| Kernel | Skylake | ThunderX2 | Zen 2 |
| --- | --- | --- | --- |
| CPML4_X | 0.185s (rank 0) | 0.160s (rank 15) | 0.0883s (rank 15) |
| CPML4_Y | 0.185s (rank 0) | 0.162s (rank 3) | 0.0896s (rank 15) |
| CPML4_Z | 0.185s (rank 3) | 0.165s (rank 15) | 0.0905s (rank 15) |

The results show that the platforms compute the three CPML4 at similar times, especially the Skylake. The ranks 0, 3, and 15 process corners, which are the most imbalanced regions for most platforms, as seen in Section IV-C. ThunderX2 is the most time-consuming platform for the Ondes3D entire execution. Still, it processes CPML4 faster than the Intel most-recent architecture evaluated, which indicates that ThunderX2 imbalance is unrelated to this computation. Zen 2 presents the best results, with rank 15 computing more load for all cases, with values less than half the Skylake times.

## V. RELATED WORK

Previous works that use a varied set of multicore platforms usually focus on evaluating the platforms' specificities or using them to support new findings [17], [18], [19]. Scientific applications with diverse memory access and communication characteristics, for example, can benefit from these platforms.

Evaluations of architectures from different vendors measure aspects such as the power consumption and the energy-saving, the memory access intensity, and the workload distribution on heterogeneous architectures [20], [21], [22]. In the end, the challenge of selecting the best platform remains to the programmer since the performance depends on the application's specifications and scalability limitations [22], [23].

Some approaches evaluate a specific application across different multicore platforms. A paper investigates the performance of a sparse matrix-vector multiply in four architectures from different vendors [24]. They conduct a detailed study on each system, correlating the performance findings with the architectural designs. They present optimization strategies, such as the potential advantages of using a multicore-specific Pthreads implementation over a traditional MPI. Another paper evaluates the performance of a parallel astrophysics application for four Intel multiprocessors using MPI [25]. They characterize the load imbalance resultant from an adaptive mesh refinement using Integrated Performance Monitoring for software instrumentation. Their work only gathers data related to MPI operations, focusing on scalability and communication behaviors for a global execution evaluation.

As far as we know, none work measures the multicore architecture's impact on the temporal load imbalance of a scientific application using its execution signature. Also, none explored a varied multicore architecture set with eight different generations and vendors and performed evaluations about the kernel's execution.

## VI. DISCUSSION AND CONCLUSION

Proper load distribution is key in scientific parallel applications. The Ondes3D seismic wave simulator has a spatial and temporal load imbalance. We evaluate the Ondes3D temporal imbalance on eight different microarchitectures for the application's macro kernels in two earthquake scenarios.

We conclude that the temporal load imbalance of Ondes3D can be minimized when exploring different architectures. The AMD Zen 2 is the best choice considering execution time, but most importantly, the spatial imbalance for both scenarios. This architecture computes all kernels faster than other platforms but still presents consistent temporal imbalance. The ARM ThunderX2 is the worst choice considering the execution time, with the highest spatial imbalance and the less temporal imbalance. One rank of the Chuetsu-Oki execution at Zen 2 computes up to 5 times faster than at ThunderX2.

Our findings also reveal that the AMD Zen presents similar load distribution to the Intel processors for Chuetsu-Oki. For a more complex workload, it shows more similarities with AMD Zen 2, its following architecture. Our macro kernels evaluation shows that Chuetsu-Oki presents a temporal imbalance on the borders due to the Absorbing Conditions for all architectures except ThunderX2. For Sichuan, the temporal imbalance covers almost all the geological domain for the Intel architectures, while AMD and ARM executions present only subtle temporal imbalance. Our microkernels investigation shows that ThunderX2 is faster than Intel, leaving an open research issue for future studies on its imbalance.

A temporal load imbalance evaluation can impact load balancing decisions on scientific applications. A previous work applied a dynamic load rebalancing to overcome the Ondes3D temporal imbalance [7]. However, for cases where the spatial imbalance is stronger, as happens in ThunderX2, a rebalancing approach could limit itself to measure the imbalance initially and assemble a new load distribution that lasts until the end of the execution. As future work, we intend to investigate the temporal load imbalance on fine-grain topologies. We also want to investigate the numerical results of executions on different architectures to identify how they impact the Ondes3D results.

**Software and Data Availability**. We endeavor to make our analysis reproducible. A public companion hosted at https:

//gitlab.com/anaveroneze/hpcs2020 contains the source code, datasets, and instructions to reproduce our results. A perennial archive is also available in Zenodo at https://zenodo.org/record/4551585.

ACKNOWLEDGEMENTS

This study was financed in part by the Coordenação de Aperfeiçoamento de Pessoal de Nível Superior - Brasil (CAPES) - Finance Code 001, under grant no 88887.481194/2020-00. The experiments were executed on the PCAD at the Federal University of Rio Grande do Sul, and on Grid'5000, supported by Inria, CNRS, RENATER and other organizations.